\documentclass[epj]{webofc}
\usepackage[varg]{txfonts}   
\usepackage{graphicx}
\usepackage{subfigure}
\usepackage{pstricks}
\usepackage{psfrag}
\usepackage{multirow}
\usepackage{pifont}

\newcommand{\sz}{\ensuremath{\sigma^z}}
\renewcommand{\sp}{\ensuremath{\sigma^+}}
\newcommand{\sm}{\ensuremath{\sigma^-}}
\def\Id{\mathbb{1}}
\newcommand{\X}{\ding{53}}

\wocname{EPJ Web of Conferences}
\woctitle{CONF12}
\begin{document}
\selectlanguage{english}
\title{{\vspace{-15mm} \normalsize\hfill{\small DESY 16-212}}\\[10mm]
Towards overcoming the Monte Carlo sign problem\\ with tensor networks}
%
%

\author{Mari Carmen Ba\~nuls\inst{1} \and
        Krzysztof Cichy\inst{2,3}\thanks{\email{kcichy@th.physik.uni-frankfurt.de}} \and
        J.~Ignacio Cirac\inst{1} \and
        Karl Jansen \inst{4} \and
        Stefan K\"uhn \inst{1} \and
        Hana Saito \inst{5}
}

\institute{Max-Planck-Institut f\"ur Quantenoptik, Hans-Kopfermann-Straße 1, 85748 Garching, Germany
\and
           Goethe-Universit\"at Frankfurt am Main, Institut f\"ur Theoretische Physik,
Max-von-Laue-Stra{\ss}e 1, 60438 Frankfurt am Main, Germany
\and
           Adam Mickiewicz University, Faculty of Physics, Umultowska 85, 61-614 Pozna\'n, Poland
\and
NIC, DESY, Platanenallee 6, 15738 Zeuthen, Germany
\and
AISIN AW Co., Ltd., 10 Takane, Fujii-cho, Anjo, Aichi, 444-1192, Japan
}

\abstract{%
  The study of lattice gauge theories with Monte Carlo simulations is hindered by the infamous sign problem that appears under certain circumstances, in particular at non-zero chemical potential. So far, there is no universal method to overcome this problem. However, recent years brought a new class of non-perturbative Hamiltonian techniques named tensor networks, where the sign problem is absent. In previous work, we have demonstrated that this approach, in particular matrix product states in 1+1 dimensions, can be used to perform precise calculations in a lattice gauge theory, the massless and massive Schwinger model. We have computed the mass spectrum of this theory, its thermal properties and real-time dynamics. In this work, we review these results and we extend our calculations to the case of two flavours and non-zero chemical potential. We are able to reliably reproduce known analytical results for this model, thus demonstrating that tensor networks can tackle the sign problem of a lattice gauge theory at finite density.
}
\maketitle
\section{Introduction}
\label{intro}
Lattice methods have proven to be indispensable when addressing questions in the non-perturbative regime of quantum chromodynamics (QCD).
Lattice QCD (LQCD), initially introduced in the 1970s by Wilson~\cite{Wilson1974}, has succeeded in elucidating many aspects of the strong interaction from first principles, including e.g.\ the computation of quark and hadron masses, hadron structure related quantities as well as non-zero temperature properties.
The standard tool of LQCD are Monte Carlo (MC) simulations, performed on world's most powerful supercomputers.
Despite the great progress achieved in the last decades, there are still areas where LQCD fails to give precise quantitative answers.
This concerns, in particular, parameter regimes where MC simulations encounter a sign problem.
Arguably the most important such case, at least from the point of view of this paper, is at non-zero chemical potential, i.e.\ at finite baryon density.
When the chemical potential is non-zero, the standard Boltzmann factor of LQCD in the Euclidean formulation becomes complex and can no longer be interpreted as a probability measure.
This undermines the whole principle of MC simulations.
Although partial solutions to this problem are known, they only allow to simulate relatively small baryon densities, in particular much smaller than the one corresponding to the conjectured critical endpoint in the density-temperature phase diagram.
As a consequence, there is broad search for alternative approaches, including Lefschetz thimbles \cite{Scorzato:2015qts}, complex Langevin simulations \cite{Sexty:2014zya} and density of states methods \cite{Langfeld:2016kty}.
Yet another thread of research, one that we concentrate on in this paper, is related to the tensor networks (TN) approach.

The TN approach\footnote{For a pedagogical introduction, see e.g.\ Refs.~\cite{Verstraete2008,Orus:2013kga}.}, originally introduced in the context of condensed matter physics and further developed thanks to quantum information theory, has been successfully applied to the description of quantum many-body systems, including, in last years, also lattice field theory.
The latter included computations of spectra~\cite{Banuls:2013jaa,Buyens:2013yza,Buyens:2014pga,Buyens:2015dkc}, thermal states~\cite{Banuls:2015sta,Banuls:2016lkq,Saito:2014bda,Saito:2015ryj,Buyens:2016ecr}, phase diagrams~\cite{Silvi2016,Zohar:2015eda,Zohar:2016wcf,Tagliacozzo:2012vg} and real-time evolution~\cite{Buyens:2013yza,Kuhn:2015zqa,Pichler:2015yqa,Buyens:2015tea}.
In this paper, we review the approach and our results for the Schwinger model.
Moreover, we go one step further by simulating the two-flavour Schwinger model in a setup where MC simulations would suffer from the sign problem, at non-zero chemical potential.

The paper is organized as follows.
We formulate the Schwinger model for a general number of flavours in Sec.~2.
In Sec.~3, we describe the basics of the tensor network method.
Sec.~4 summarizes our findings for the chiral condensate in the one-flavour model at zero temperature.
In Sec.~5, we review our results at finite temperature, considering again the chiral condensate.
In Sec.~6, we discuss the results from the two-flavour Schwinger model at non-zero chemical potential.
Finally, Sec.~7 summarizes and a short discussion of the prospects of the TN approach for lattice gauge theories is given.

\section{Multi-flavour Schwinger model}
\label{sec-2}
The Schwinger model \cite{Schwinger:1962tp} is quantum electrodynamics in 1+1 dimensions (QED$_2$).
It has been widely used as a toy model for testing new lattice methods, before applying them e.g.\ to QCD.
Interestingly, QED$_2$ shares certain properties with QCD$_4$, such as confinement and chiral symmetry breaking.

Our lattice formulation uses Kogut-Susskind staggered fermions~\cite{Kogut1975}, leading to the following Hamiltonian for $N$ sites and $F$ flavours, with open boundary conditions (OBC):
 \begin{align}
 \begin{aligned}
 H = &-\frac{i}{2a}\sum_{n=0}^{N-2}\sum_{f=0}^{F-1}\left(\phi^\dagger_{n,f}e^{i\theta_n}\phi_{n+1,f}-\mathrm{H.c.}\right)
 +\sum_{n=0}^{N-1}\sum_{f=0}^{F-1}\left(m_f(-1)^n +\kappa_f \right)\phi^\dagger_{n,f}\phi_{n,f}
 + \frac{ag^2}{2}\sum_n L_n^2,
 \end{aligned}
 \label{hamiltonian}
\end{align}
where: $a$ -- lattice spacing, $g$ -- coupling, $m_f$ -- $f$-th flavour fermion mass, $\kappa_f$ -- its chemical potential, $\phi_{n,f}$ -- fermionic fields describing flavour $f$ on site $n$, $L_n$ -- operator that gives the electric flux of link $n$, $\theta_n\in[0,2\pi]$ -- conjugate operator such that $e^{\pm i\theta_n}$ acts as the electric flux raising/lowering operator.
The Gauss law is expressed by:
\begin{equation}
L_n = L_{n-1} + \sum_{f=0}^{F-1}\left( \phi^\dagger_{n,f}\phi_{n,f}-\frac{1}{2}(1-(-1)^n)\right)
\end{equation}
and allows to integrate out the gauge fields, leaving only the electric field on the left boundary as an independent parameter (taken to be zero in this work).

We write the above Hamiltonian in dimensionless form and we use a Jordan-Wigner transformation, $\phi_n = \sum_{l<n}(i\sz_l)\sm_n$, $\phi_n^\dagger = \sum_{l<n}(-i\sz_l)\sp_n$, 
where $\sigma_n^{z/\pm}$ are Pauli matrices acting on site $n$ and where we order the fermions on each site such that $\phi_{n,f} = \phi_{nF+f}$. 
The Hamiltonian in this equivalent spin formulation describes a spin chain of length $NF$ and reads:
\begin{align}
\begin{aligned}
 W=&-x\sum_{p=0}^{NF-1}\left(\sp_{p}(i\sz_{p+1})\dots (i\sz_{p+F-1})\sm_{p+F} + \mathrm{h.c.}\right)
 +\sum_{n=0}^{N-1}\sum_{f=0}^{F-1}\bigl(\mu_f(-1)^n +\nu_f \bigr)\frac{1+\sz_{nF+f}}{2}\\
 &+\sum_{n=0}^{N-2}\left(\frac{F}{2}\sum_{k=0}^n(-1)^k + \frac{1}{2}\sum_{k=0}^n\sum_{f=0}^{F-1}\sz_{kF+f}\right)^2,
\end{aligned}
\label{spin_hamiltonian_raw}
\end{align}
where $x=1/(ag)^2$, $\mu_f = 2\sqrt{x}m_f/g$, $\nu_f = 2\sqrt{x}\kappa_f/g$ are dimensionless parameters.
Note that the gauge symmetry (manifested in the Gauss law) has effectively generated long-range interactions, of the form $\sigma^z_n \sigma^z_m$ (with arbitrary site-flavour index $n,m$) in the spin language, and also a site-dependent magnetic field. The hopping term always involves fermions of the same flavour, i.e.\ it acts on pairs of spins at a distance $F$ in the effective $NF$-site chain.

In the following, we consider the 1-flavour and 2-flavour cases.
With the former, one can demonstrate, for example, calculations of the spectrum, of ground state expectation values and of thermal properties.
However, the chemical potential has no physical effect with only one flavour -- therefore, to exhibit that the TN approach can tackle the model at non-zero density, we employ the 2-flavour case.
In this way, we show that in the regime where a MC simulation would suffer from a sign problem, the TN methods can still produce reliable results.

\section{Tensor network approach}
\label{sec-3}
Among the tensor network methods, the most successful one for one-dimensional systems (1+1-dimensional quantum field theories) is the matrix product states (MPS) \cite{aklt88,kluemper91,kluemper92,fannes92fcs,Verstraete:2004zza,perez07mps}.
Below, we use MPS for all our investigations at zero temperature.
The MPS ansatz for a system of $N$ sites has the form:
\begin{equation}
 |\Psi\rangle =\sum_{i_0,\ldots i_{N-1}=0}^{d-1} {\rm Tr}\, (A_0^{i_0}\ldots A_{N-1}^{i_{N-1}}) |i_0\ldots i_{N-1}
\rangle,
\end{equation}
 where $\{|i_n\rangle\}_{i=0}^{d-1}$ are basis states for each site $n$ and $d$ is the dimension of the one-site Hilbert space.
Each matrix $A_n^{i_n}$ is $D$-dimensional and $D$ is called the bond dimension and determines the number of parameters in the MPS ansatz.
It has been shown that MPS describe accurately ground states of local gapped Hamiltonians, but their practical application extends to more general models.

\begin{figure}[t!]
\centering
\psfrag{P}[c][l]{$|\Psi\rangle$}
\psfrag{A}[c][l]{$A_k$}
\psfrag{H}[c][c]{$\langle\Psi|H|\Psi\rangle$}
\psfrag{N}[c][l]{$\langle\Psi|\Psi\rangle$}
\includegraphics[width=.65\columnwidth]{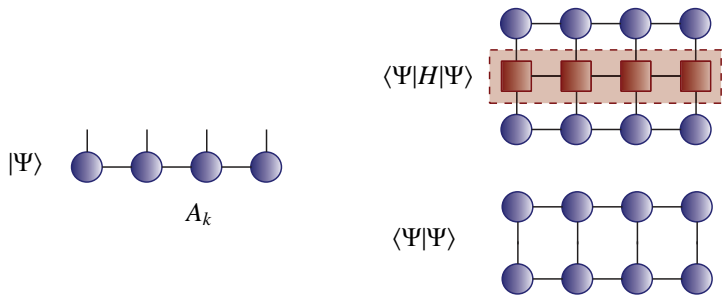}
\caption{
The commonly used pictorial representation of MPS. See text for description.
}
\label{fig:TN}
\end{figure}

The task of finding the MPS approximation to the ground state (GS) boils down to finding the components of the tensors $A_n$ that minimize the expectation value of the Hamiltonian, $E=\langle\Psi|H|\Psi\rangle/\langle\Psi|\Psi\rangle$.
This is performed variationally, by minimizing $E$ with respect to one tensor $A_n$ at a time and changing $n$ successively (sweeping from left to right and back) until global convergence is achieved.
Such local minimization problem consists in solving the eigenvalue problem of an effective Hamiltonian acting on site $n$ and its neigbouring virtual bonds.
This variational algorithm \cite{Verstraete:2004zza} is related to the original density matrix renormalization group (DMRG) formulation \cite{white92dmrg,schollwoeck05dmrg}, which was better understood in the language of MPS.
Note that we use OBC, which is, in general, more efficient and results in more stable numerical behaviour.
The commonly used graphical representation of an MPS and its contractions that give the energy $E$ is shown in Fig.~\ref{fig:TN}.
A circle corresponds to one tensor $A_n$ and the legs represent indices (vertical ones $0,\ldots,d-1$; horizontal ones $0,\ldots,D-1$).
The lines connecting two tensors represent contracted indices, while open legs of an MPS correspond to physical indices of each site.
The contraction of an MPS $|\Psi\rangle$ with itself (contracting the physical indices) represents the norm $\langle\Psi|\Psi\rangle$ (lower right of Fig.~\ref{fig:TN}), while an insertion of an operator, such as the Hamiltonian, allows to compute its expectation value in the state represented by an MPS, e.g.\ $\langle\Psi|H|\Psi\rangle$ (upper right).

Apart from the GS, the variational algorithm can find excited states.
Knowing the ground state MPS, $|\Psi_0\rangle$, this proceeds by constructing a projector onto the orthogonal subspace and finding the GS in this subspace.
Successively projecting out further excited states, one can find any number of them (however, the accumulated error may become large for higher excited states).
In this way, one can access the vector and scalar particles in the Schwinger model spectrum.
Note, however, that the usage of OBC violates translational invariance and hence identification of particles, which are zero-momentum excitations, is more difficult.
In particular, the scalar meson, naively the second excitation of the one-flavour Schwinger model, is above momentum excitations of the vector particle.
Nevertheless, unambiguous identification of particles is still possible.

The application of the MPS approach to non-zero temperatures requires its extension allowing for a description of operators, in this case density matrix operators~\cite{verstraete04mpdo,zwolak04mpo,pirvu10mpo}.
One introduces matrix product operators (MPO) of the form:
\begin{equation}
   \rho
   = 
         \sum_{\{i_n,j_n\}}{\rm Tr} \left( M_0^{i_0j_0} \cdots M_{N-1}^{i_{N-1}j_{N-1}} \right) |i_0\ldots i_{N-1}\rangle\langle j_0\ldots j_{N-1}|,
         \label{eq:MPO}
\end{equation}
where the tensors $M_n$ play an analogous role to the tensors $A_n$ in the MPS ansatz.
For density matrices appropriate for the description of a system at finite temperature, an MPO approximation can be obtained using imaginary time evolution of the identity operator~\cite{verstraete04mpdo}, 
$\rho(\beta)\propto e^{-\beta H}=e^{-\frac{\beta}{2} H} \Id e^{-\frac{\beta}{2} H}$, with $\beta\equiv1/T$ being the inverse temperature.
For a detailed description of the technical steps for the one-flavour Schwinger model, we refer to our original papers~\cite{Banuls:2015sta,Banuls:2016lkq}.
The efficient application of the imaginary time evolution to find thermal states requires specific algorithms, for instance applying discrete time steps in a Suzuki-Trotter expansion \cite{trotter59,suzuki90} plus using appropriate approximations for the long-range terms in the Hamiltonian.
We start with the identity operator, $\rho(0)$, which can be represented trivially by an MPO with bond dimension one.
Then, we evolve the density matrix and approximate each step by an MPO with the desired maximum bond dimension, using the so-called Choi isomorphism~\cite{choi}, $|i\rangle\langle j| \rightarrow |i\rangle\otimes |j\rangle$ that vectorizes the density operators, i.e.\ transforms the MPO into an MPS with physical dimension $d^2$.
Finally, we minimize the Euclidean distance between the original and final MPS and repeat the procedure until reaching the inverse temperature $\beta$.

\section{Results -- 1-flavour Schwinger model at $T=0$}
\label{sec-4}
We start reviewing our results with a zero-temperature calculation of the ground state in the one-flavour Schwinger model.
We illustrate the main steps concerning an example of the determination of the ground state expectation value of the chiral condensate, $\Sigma=\left\langle {\bar \psi}\psi \right\rangle$, 
which can be written using spin operators as: 
\begin{equation}
 \frac{\Sigma}{g} = \frac{\sqrt{x}}{N} \sum_n (-1)^n \frac{1+\sigma_n^z}{2}.
\end{equation}
In the massless case, this expectation value can be computed analytically, yielding the result $\Sigma_{T=0,m=0}/g=e^{\gamma_E}/2\pi^{3/2}\approx0.159929$~\cite{Sachs:1991en}, where $\gamma_E\approx0.577216$ is the Euler-Mascheroni constant.
The analytical computation at non-zero fermion mass is no longer possible.
The non-perturbative MPS determination requires subtraction of a logarithmic divergence, which is present already in the free theory and can be removed with:
\begin{equation}
\label{eq:subtr}
\Sigma_{\rm subtr}(m/g,x)=\Sigma(m/g,x)-\Sigma_{\mathrm{free}}(m/g,x),
\end{equation}
where $\Sigma_{\mathrm{free}}(m/g,x)$ is the free theory condensate at fermion mass $m/g$ and lattice spacing corresponding to the inverse coupling $x$.

\begin{figure}[t!]
\centering
\includegraphics[width=0.345\textwidth,angle=270]{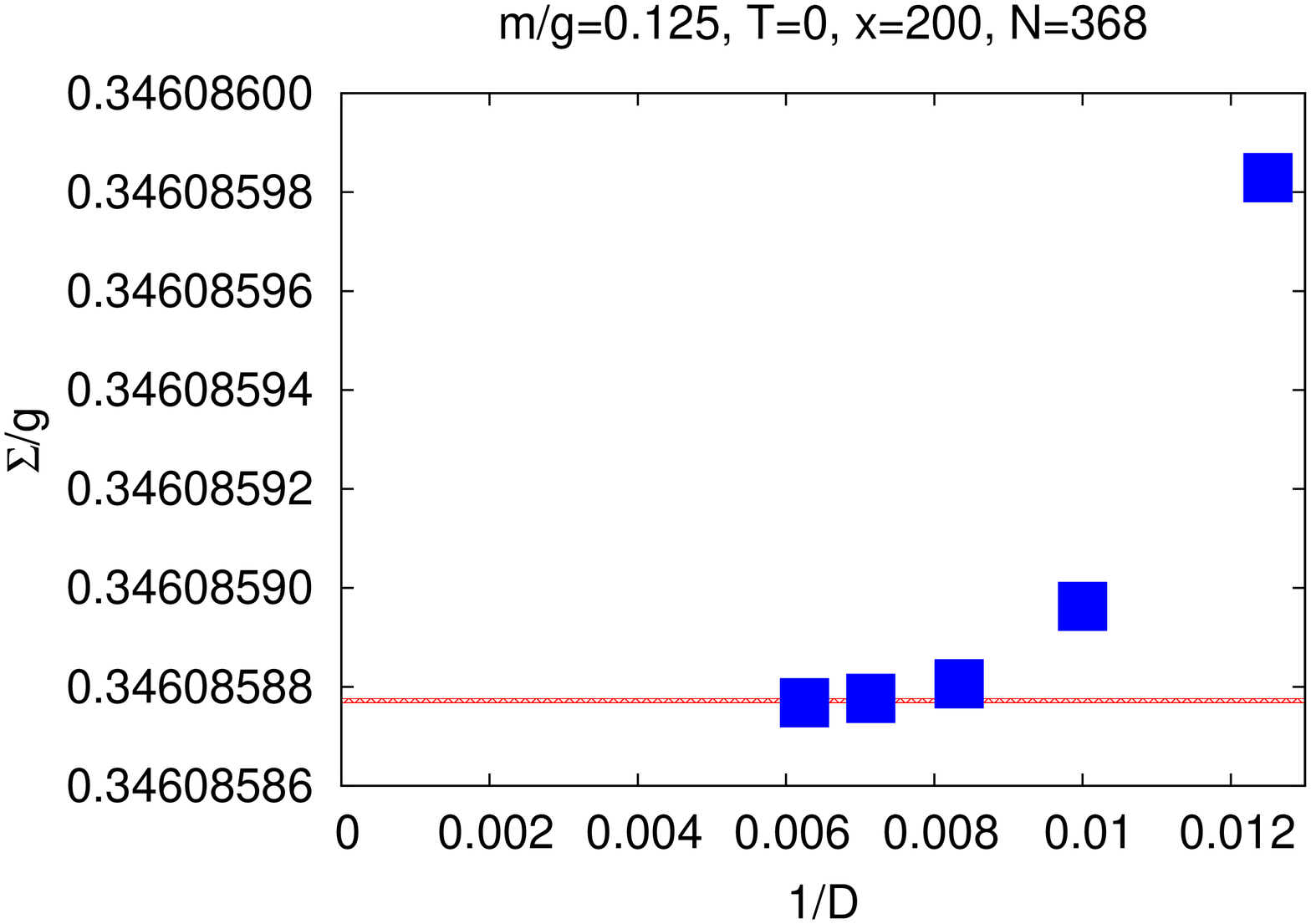}
\includegraphics[width=0.345\textwidth,angle=270]{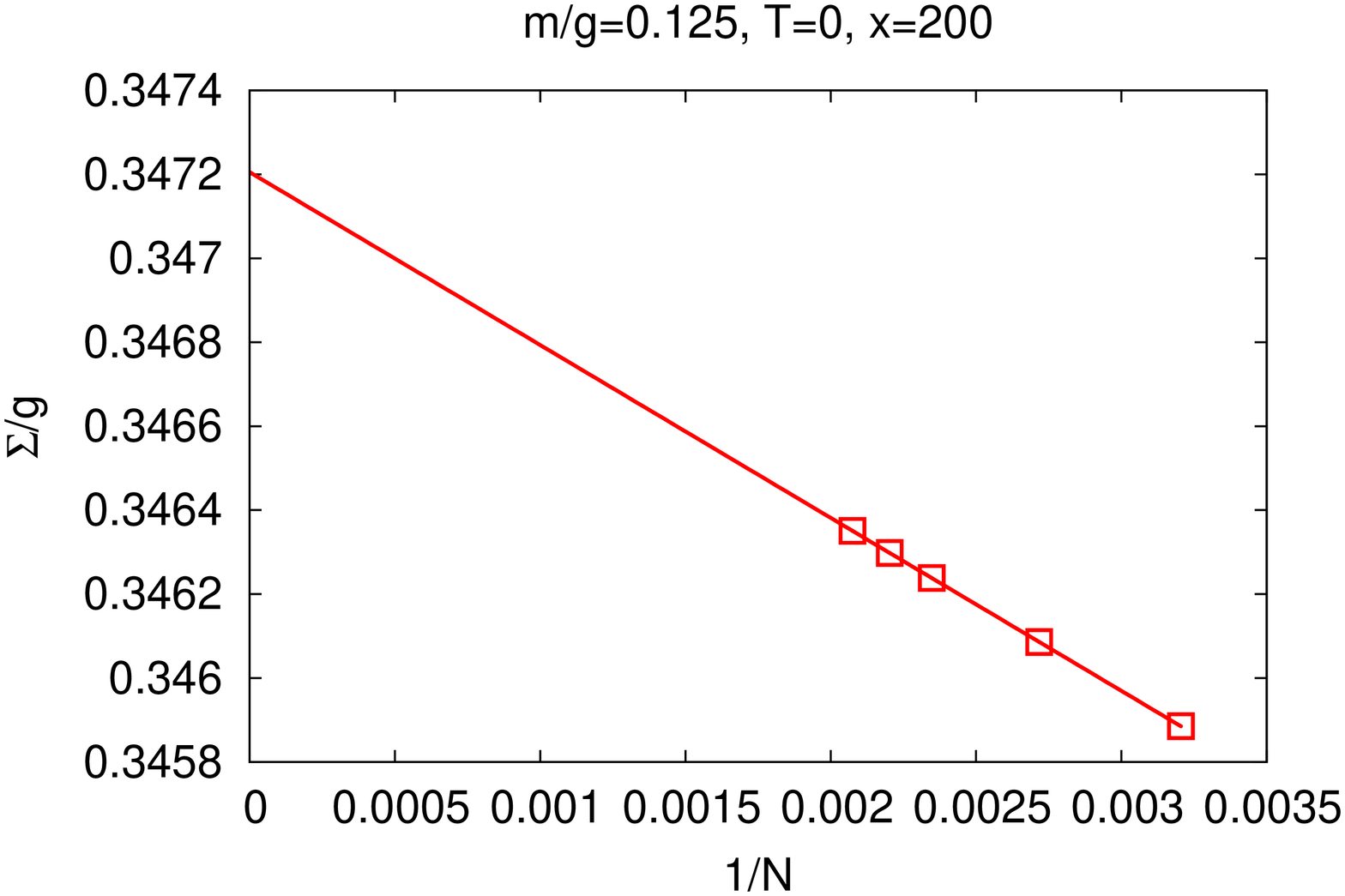}
\includegraphics[width=0.345\textwidth,angle=270]{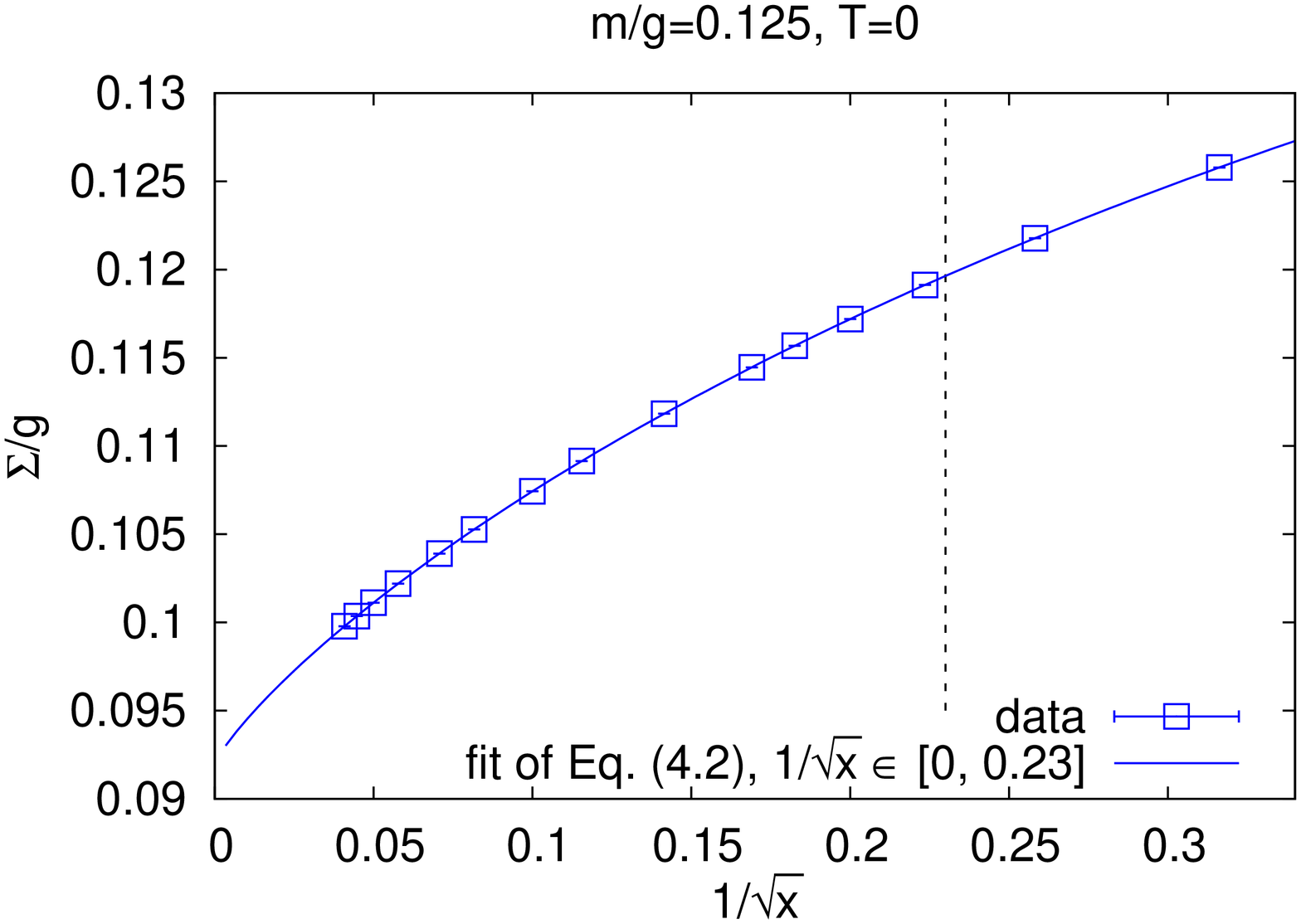}
\caption{
Successive extrapolations needed to obtain the continuum result for the $T=0$ chiral condensate at fermion mass $m/g=0.125$.
Upper left: extrapolation in the inverse bond dimension ($x=200$, $N=368$). Upper right: extrapolation to infinite volume ($x=200$).
Lower: continuum extrapolation.
}
\label{fig:cond}
\end{figure}

An MPS simulation is performed for a finite bond dimension $D$, with a chain of finite length $N$ and at a finite lattice spacing, given by the parameter $x$.
In Fig.~\ref{fig:cond}, we illustrate the sequence of extrapolations needed to obtain the continuum result, which is the aim of the computation.
First, we extrapolate to infinite bond dimension, thus removing the effect of approximating the ground state with an MPS of finite $D$.
The typical behaviour, illustrated in the upper left plot of Fig.~\ref{fig:cond}, is saturation of finite bond dimension effects at rather moderate value of $D$ of order 100.
As our final value, we take the one corresponding to the maximum computed bond dimension, $D=160$ in this case, and we estimate its uncertainty to be of order of the difference between the results at $D=160$ and $D=140$.
Note that, given the approximately exponential approach to the value at $1/D=0$ that we observe in many cases, this error estimate is rather conservative.
It should be emphasized that the central value obtained in this way is always compatible with the result of exact diagonalization (ED), in case the latter is feasible.\footnote{ED is possible for spin systems including up to around 20-24 spins with a naive approach and up to 40-48 spins when symmetries of the problem are fully exploited. For $N=368$, the exponential complexity of the problem renders the ED computation practically impossible, as the dimension of the Hamiltonian matrix is of $\mathcal{O}(10^{110})$.}
Having the estimates of the $1/D=0$ values of the condensate for a few values of the volume $N$, one can perform the infinite volume extrapolation, as illustrated in Fig.~\ref{fig:cond} (upper right).
In order that the leading order $1/N$ behaviour is reliable, one needs to ensure that the ratio $N/\sqrt{x}$ is large enough, in this case of order 20-30.
Finally, one can subtract the free theory condensate (Eq.~(\ref{eq:subtr})) and perform the continuum extrapolation.
In the illustrated case (lower plot of Fig.~\ref{fig:cond}), we use Eq.~(4.2) of Ref.~\cite{Banuls:2016lkq}, which is a fitting ansatz quadratic in the lattice spacing $1/\sqrt{x}$ and including logarithmic corrections of $\mathcal{O}(\log(x)/\sqrt{x})$.
Note that we also test sensitivity to higher order corrections of $\mathcal{O}(1/x^{3/2})$ and to the choice of the fitting interval in $x$.
In the end, we obtain a continuum result with all sources of possible uncertainties fully estimated and quantified.

\begin{table}[t!]
\begin{center}
\begin{tabular}{|c|c|c|c|}
\hline
&
\multicolumn{3}{|c|}{Subtracted condensate} \\
\hline
\multirow{2}*{$m/g$}  & Our result & Ref.~\cite{Buyens:2014pga} & Exact ($m=0$) \\
&MPS & MPS& or Hosotani ($m>0$)~\cite{Hosotani:1998za}\\
\hline
0 & 0.159929(7) & 0.159929(1) & 0.159929 \\
\hline
0.0625 & 0.1139657(8) & -- & 0.1314 \\
\hline
0.125 & 0.0920205(5) & 0.092019(2) & 0.1088 \\
\hline
0.25 &  0.0666457(3) & 0.066647(4) & 0.0775 \\
\hline
0.5 & 0.0423492(20) & 0.042349(2) & 0.0464 \\
\hline
1.0 & 0.0238535(28) & 0.023851(8) & 0.0247\\
\hline
\end{tabular}
\caption{\label{tab:cond} Continuum values of the $T=0$ chiral condensate $\Sigma/g$ for different fermion masses. We compare with results of Ref.~\cite{Buyens:2014pga} and with the analytical result in the massless case or the approximated result from Ref.~\cite{Hosotani:1998za} in the massive case.}
\end{center}
\end{table}

Our final results are given in Tab.~\ref{tab:cond}, together with a comparison to another recent MPS computation~\cite{Buyens:2014pga} and the exact result ($m=0$) or an approximate result from Ref.~\cite{Hosotani:1998za} ($m>0$).
Note that the former comparison is highly non-trivial, since the authors of Ref.~\cite{Buyens:2014pga} used a different approach of infinite MPS, which works directly in the thermodynamic limit, instead of our approach of extrapolating to this limit from selected values of $N$.
In general, the obtained precision is, in both cases, very good, at the level of around five-six significant digits.

Apart from the GS expectation value of the condensate, we also computed the spectrum of the one-flavour Schwinger model, reaching similar precision.
For details, we refer to Ref.~\cite{Banuls:2013jaa}.

\section{Results -- 1-flavour Schwinger model at $T>0$}
\label{sec-5}
As described above, the thermal computation is somewhat more complicated.
However, the necessary steps in the numerical procedure are very similar, with the addition of another extrapolation that has to be performed.
Since the imaginary time interval $[0,\beta]$ is divided into discrete steps of length $\delta$, one needs to take the limit $\delta\rightarrow0$ to eliminate the error necessarily introduced by non-zero $\delta$.
This extrapolation is performed as the second one, after the bond dimension extrapolation (see Ref.~\cite{Banuls:2016lkq} for explicit plots).

\begin{figure}[t!]
\begin{center}
\includegraphics[width=.45\columnwidth]{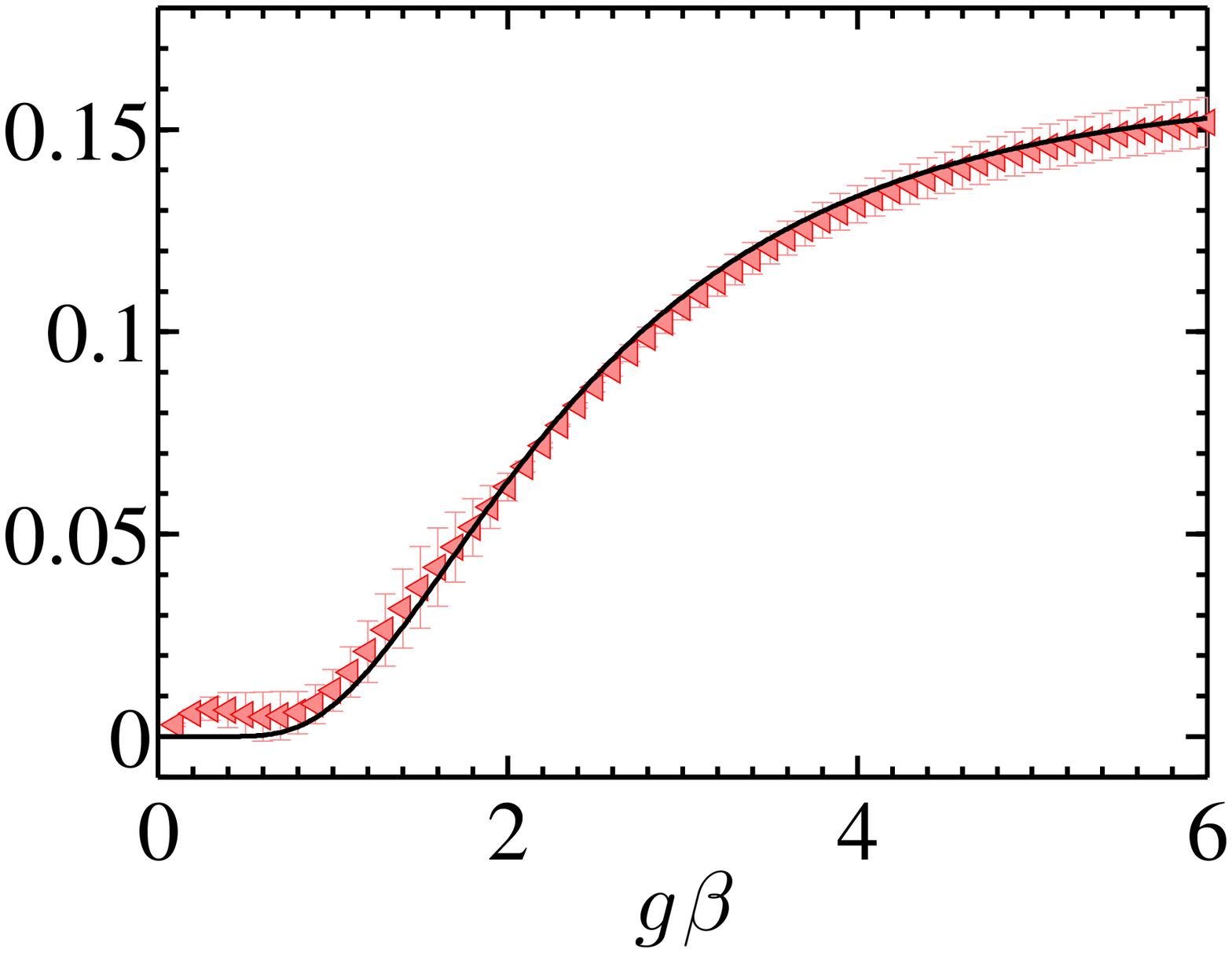}
\includegraphics[width=.45\columnwidth]{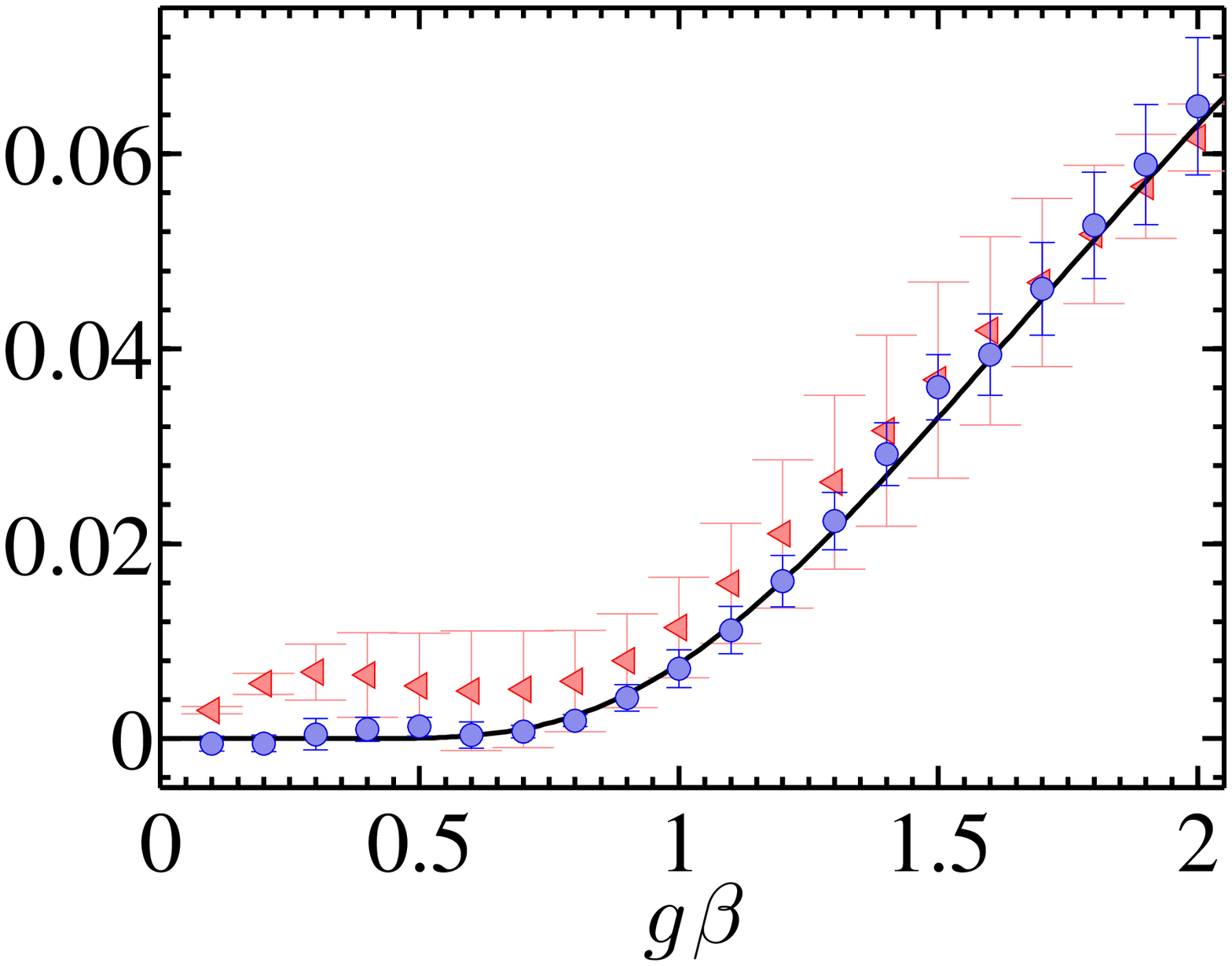}
\end{center}
\caption{\label{fig:m0} Dependence of the chiral condensate on the inverse temperature $\beta$ in the massless case. The black solid line is the analytical result of Ref.~\cite{Sachs:1991en}.
Left: the red triangles correspond to our continuum results obtained without flux truncation.
Right: in addition to the red triangles representing the results without flux truncation, we plot also the result with flux truncation applied (blue circles).}
\end{figure}

The continuum results for the massless condensate are shown in Fig.~\ref{fig:m0}.
The long-range terms in the Hamiltonian require specific approximations to implement the corresponding steps of the evolution.
The left plot shows the result when a Taylor expansion is used.
We reproduced the analytical curve of Ref.~\cite{Sachs:1991en} in a wide range of temperatures.
Note, however, that the high temperature region with $g\beta\leq0.5$ does not agree with the analytical result in this plot.
This fact, resulting from enhanced cut-off effects in this region, required further investigation.
It is possible to increase the precision of the computation by applying instead a truncation to the evolution operator, which turns out to be equivalent to a truncated model, with maximum flux on the link set to a specified value $L_{\rm cut}$. 
We found that $L_{\rm cut}=10$ effectively corresponds to the untruncated model, but the presence of the flux cut-off leads to a reduced computational effort.
Thus, we could simulate at smaller lattice spacings and obtain more reliable continuum results, illustrated in the right plot of Fig.~\ref{fig:m0}.
In addition to reducing the uncertainties in the whole range of temperatures, we obtained agreement with the analytical curve even for small $g\beta$.

\begin{figure}[t!]
\begin{center}
\includegraphics[width=0.33\textwidth,angle=270]{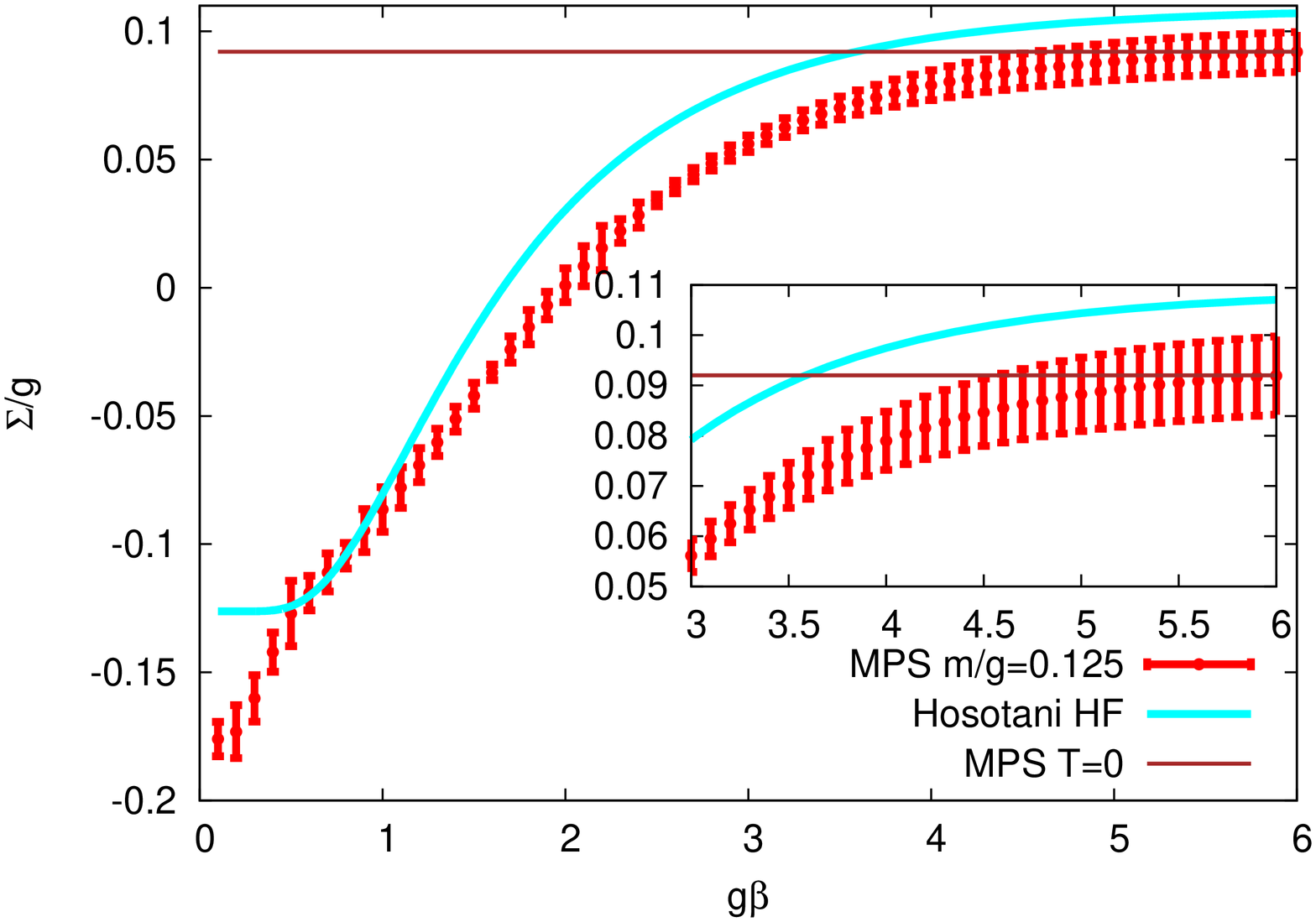}
\includegraphics[width=0.33\textwidth,angle=270]{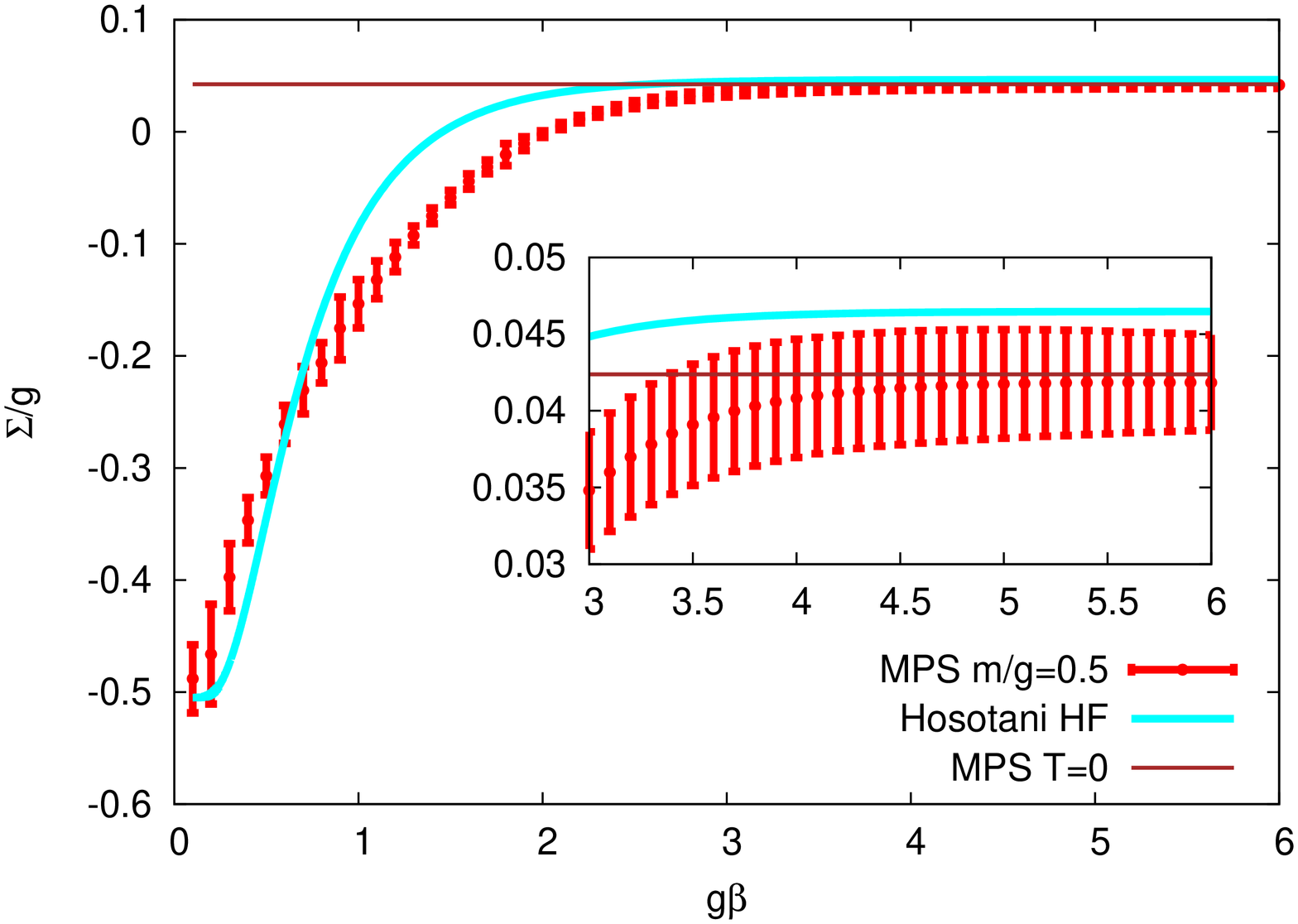}
\end{center}
\caption{\label{fig:m} Dependence of the chiral condensate on the inverse temperature $\beta$ for two selected fermion masses, $m/g=0.125$ (left) and $m/g=0.5$ (right).
Our continuum results (with flux truncation) are shown with red data points.
The cyan solid curve is the reference curve from Ref.~\cite{Hosotani:1998za} (generalized Hartree-Fock approximation) and the brown horizontal line is our continuum result for $T=0$.
}
\end{figure}

In Fig.~\ref{fig:m}, we also show our continuum result for two selected fermion masses, $m/g=0.125$ (left) and $m/g=0.5$ (right).
We note that, as expected, the thermal curves approach the $T=0$ result when $g\beta$ is increased.
The agreement with the approximate result of Ref.~\cite{Hosotani:1998za} is moderate.
However, after our work, a cross-check was provided by the authors of Ref.~\cite{Buyens:2016ecr} using the infinite MPS approach and full agreement was observed.

\section{Results -- 2-flavour Schwinger model at $T=0$ and non-zero chemical potential}
\label{sec-6}
The essential motivation for the TN approach comes from the possibility of overcoming the notorious sign problem that plagues MC simulations of QCD at non-zero chemical potential.
The results presented in the previous sections have demonstrated the feasibility of the MPS/MPO approach when applied to lattice gauge theories.
However, the analyzed system, the one-flavour Schwinger model both at zero and non-zero temperatures, does not suffer from the sign problem when simulated with MC.
Now, we move on to a case where the sign problem indeed appears -- the two-flavour model at non-zero chemical potential.
For a more detailed description of our results, see Refs.~\cite{Banuls:2016gid,Banuls:2016hhv}.

\begin{figure}[t!]
\centering
\sidecaption
\includegraphics[width=0.5\textwidth]{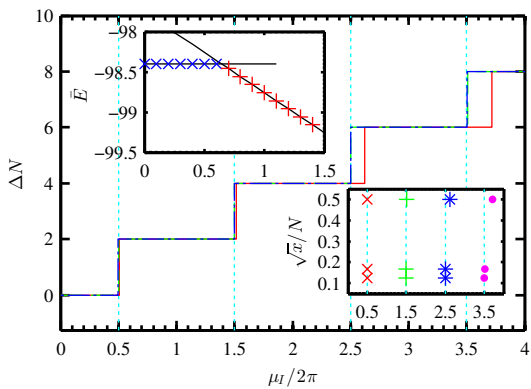}
\caption{\label{fig:jumps}
Continuum estimate for the isospin number, $\Delta N$, as a function of $\mu_I/2\pi$ ($m=0$), for volumes $N/\sqrt{x}=2$ (red solid line), 6 (green dashed line) and 8 (blue dash-dotted line). The dotted vertical lines are theoretical predictions for the phase transitions.
Upper inset: GS energy as a function of $\mu_I/2\pi$ for $N/\sqrt{x}=8$, $x=16$, $D=220$ for the two phases with smallest $\Delta N$ (blue \X's indicate $\Delta N=0$ and red crosses $\Delta N=2$). The crossing of these energies determines the transition location.
Lower inset: Continuum estimate of the location of the first phase transitions vs. inverse volume $\sqrt{x}/N$ for the first (red \X's), second (green crosses), third (blue asterisks) and fourth (magenta dots) transition.}
\end{figure}

Our aim was to reproduce the analytical result of Narayanan~\cite{Narayanan:2012qf}, who computed the isospin number, $\Delta N\equiv N_0-N_1$, i.e.\ the difference between the numbers of particles of different flavours, as a function of the isospin chemical potential, $\mu_I$, which corresponds in our notation to $N(\nu_1-\nu_0)/2x$.
It was shown that $\Delta N$ undergoes jumps that correspond to first-order phase transitions.
In our MPS setup, we could calculate the GS energy of phases with different isospin numbers.
The upper inset of Fig.~\ref{fig:jumps} shows the GS energies of the phases with $\Delta N=0$ and $\Delta N=2$ and the crossing of these energies determines the location of the first transition, which corresponds to $\mu_I/2\pi=1/2$ in the continuum.
Our result is, obviously, influenced by effects of finite $D$, $N$ and $x$.
We extrapolate away these effects in a similar manner as for the chiral condensate.
In the end, we obtain transition locations as illustrated in the main plot of Fig.~\ref{fig:jumps}, provided the volume is large enough (see the lower inset for effects of too small $N/\sqrt{x}$).
Note that the result of Ref.~\cite{Narayanan:2012qf} is volume-independent.
In our case, residual finite volume effects can occur, related to the fact that $N$ is finite and thus $\Delta N$ can become of the order of $N$.
As soon as we satisfy the condition $\Delta N\ll N$, we no longer observe any dependence on the volume.
In this way, our result is in full agreement with the analytical one.
We also investigated the massive case, for which no analytical result is available.
Contrary to the zero fermion mass case, volume-independence does not hold any more, since the fermion mass provides an additional energy scale in the problem, which affects the locations of the phase transitions.

\section{Conclusion and prospects}
\label{sec-7}
Tensor networks are a poweful tool for treatment of general physical systems, especially in lower dimensions.
The research of the past few years has demonstrated that they are appropriate for the description of lattice gauge theories and they can provide precise and reliable results both at zero and  non-zero temperature.
In particular, they are free of the sign problem and hence remain as an interesting direction for future simulations of QCD at non-zero chemical potential.
For this, extension of their efficiency for higher dimensions is obviously necessary.
Theoretical progress in higher dimensional tensor networks is steadily being made.
The natural generalization of MPS to higher dimensions is Projected Entangled Pair States (PEPS) \cite{Verstraete2004b,2008PhRvL.101y0602J}.
PEPS and other related developments give good prospects of tackling systems in more than one dimension, although they are computationally much more demanding than MPS.
Currently, TN methods have already provided one of the most accurate results for certain classes of spin systems in two dimensions, especially for cases where a quantum MC simulation encounters a sign problem.
Nevertheless, the application of TN for 2+1-dimensional lattice gauge theories remains difficult.
However, it is our aim and the aim of the TN lattice gauge theory community to continue in this direction, which in the future can lead to a full application to 3+1-dimensional QCD.

\vspace*{3mm}
\noindent\textbf{Acknowledgments.} K.C.\ was supported by the Deutsche Forschungsgemeinschaft (DFG), project nr. CI 236/1-1.

\bibliographystyle{woc}
\bibliography{mps}

\end{document}